\documentstyle[twocolumn,floats,aps]{revtex}
\begin{document}
\draft
\twocolumn[\hsize\textwidth%
\columnwidth\hsize\csname@twocolumnfalse\endcsname
\title {Extended Self-similarity in Kinetic Surface Roughening}
\author{Arindam Kundagrami and Chandan Dasgupta}
\address{Department of Physics, Indian Institute of Science,\\
Bangalore -- 560 012, India.}
\author{P. Punyindu and S. Das Sarma}
\address{Department of Physics, University of Maryland,\\
College Park, MD 20742-4111.}
\date{\today}
\maketitle
\begin{abstract}
We show from simulations that a limited mobility
solid-on-solid model of kinetically rough
surface growth exhibits extended self-similarity
analogous to that found in fluid turbulence. The range over which 
scale-independent power-law behavior is observed is significantly enhanced
if two correlation functions of different order, such as 
those representing two different
moments of the difference in height between two points, 
are plotted against
each other. This behavior, found in both one and two dimensions, suggests that
the ``relative'' exponents may be more fundamental than the ``absolute'' ones.
\end{abstract}
\vskip1pc]
\narrowtext

Scale-invariant spatio-temporal behavior is observed in a wide variety
of far-from-equilibrium systems. In analogy with the 
``universality''
found in the equilibrium scaling behavior of systems near a second-order
phase transition, it is interesting to enquire about the similarities
between different non-equilibrium systems exhibiting scale-invariant behavior.
In this paper, we point out a remarkable similarity between the scaling
behavior of two well-known and extensively studied non-equilibrium systems --
turbulent fluids and growing interfaces. Krug \cite{kr1} 
discovered a similarity between the intermittent multiscaling behavior
of structure functions in strongly developed turbulence \cite{rev2,book}
and the scaling
properties of correlation functions of height fluctuations in 
simple solid-on-solid models of kinetically rough epitaxial growth \cite{dt} 
with limited surface mobility. The multiscaling
properties of these growth models have been subsequently investigated in detail
\cite{ms1} and a mechanism for this behavior has been proposed 
\cite{us1}. In this paper, we demonstrate that the
{\em extended 
self-similarity} (ESS) 
\cite{ess1} exhibited by the structure functions in fluid turbulence is
also present in the behavior of 
correlation functions of height fluctuations in these growth
models, and thereby establish
that the analogy between {\em deterministic} turbulence in fluids
and {\em stochastically driven} interface growth is remarkably deep.
We emphasize that the ESS phenomenology in our discrete stochastic
growth model is formally {\em identical} to that found in the
intermittent fluid turbulence problem, establishing a precise
one to one correspondence between these two seemingly completely
different physical processes. While the exact reasons for this 
precise analogy between these two distinct problems remain unclear
at this stage, we speculate that the existence of an infinite number
of relevant (marginal) operators in both cases may be the 
underlying mathematical cause for this analogy \cite{us1}.

We begin by pointing out
the analogy \cite{kr1} between fluid turbulence and surface growth.
In fully developed turbulence, scaling behavior is observed
in the {\em inertial range} $\eta \ll r \ll L$, where $r$ is the
length scale of interest, $L$ is the outer {\em integral scale} at which
energy is injected into the system and $\eta$
is the inner {\em dissipation scale}. A measure
of the 
separation between the inner and outer scales is provided by the {\em Reynolds
number} $Re \propto (L/\eta)^{4/3}$. 
The $q$-th order longitudinal {\em structure functions} are defined as
\begin{equation}
D_q(r)= \langle [u({\bf x} + {\bf r}, t) - u({\bf x},t)]^q \rangle,
\label{eqn1}
\end{equation}
where $u({\bf x},t)$ is the component of the 
velocity at the position $\bf x$ at time $t$ in the
direction of the relative displacement $\bf r$, 
and the brackets $\langle \ldots \rangle$ represent a spatio-temporal average. 
These structure functions are believed to exhibit power-law scaling in the
inertial range:
\begin{equation}
D_q(r) \approx r^{\zeta_q},\,\,\, \eta \ll r \ll L.
\label{eqn2}
\end{equation}
The value of $\zeta_3$ is known \cite{book} to be exactly unity.
The measured values of $\zeta_q, \, q \ne 3$ differ appreciably from the
Kolmogorov result 
\cite{book}, $ \zeta_q = q/3$, the deviation $\delta \zeta_q \equiv
\zeta_q - q/3$ being positive for $q < 3$ and negative for $q > 3$. This is
the phenomenon of {\em multiscaling} which is believed \cite{rev2,book}
to arise from the
strongly {\em intermittent} (violently fluctuating) character of the local 
energy dissipation rate
$\epsilon({\bf x},t)$ whose spatio-temporal fluctuations
may be characterized by the quantities 
$\langle \epsilon^q({\bf x},t) \rangle /\langle \epsilon({\bf x},t) 
\rangle^q$.
These quantities are expected to exhibit a power-law dependence on the Reynolds number:
\begin{equation}
\langle \epsilon^q \rangle / \langle \epsilon \rangle^q \approx
(L/\eta)^{\mu_q}\,,
\label{eqn3}
\end{equation}
where $\mu_q > 0$ for $q > 1$. 
  
In models \cite{kr1,dt,ms1,us1} 
of growing interfaces,
the role of the velocity field $u({\bf x},t)$
is played by the variable $h({\bf x},t)$ that represents the height
of a $d$-dimensional interface at point $\bf x$ at time $t$. In
these models, the inner length scale is the lattice spacing $a_0$
which is usually set to unity. 
The outer scale is set by a {\em correlation length} $\xi$ that
initially grows in time ($\xi \propto t^{1/z}$ where $z$ is the
dynamic exponent) and eventually saturates (for $t \gg L^z$ where $L$ is
the system size) to a value of the order of $L$.
Scale-independent behavior of correlation functions is observed
for length scales $1 \ll r \ll \xi$. The correlation length $\xi$,
therefore, plays the role of the Reynolds number. The height-difference
correlation functions
\begin{equation}
G_q(r,t) = \langle |h({\bf x}+{\bf r},t) - h({\bf x},t)|^q \rangle^{1/q}\,,
\label{eqn5}
\end{equation}
where $\langle \ldots \rangle$ now represents a spatial average,
are analogous \cite{fn1} to the structure functions 
defined in
Eq.(\ref{eqn1}). These
correlation functions exhibit the following scaling behavior:
\begin{equation}
G_q(r,t) \approx r^{\zeta^\prime_q},\,\,\, 1 \ll r \ll \xi(t).
\label{eqn6}
\end{equation}
The observation \cite{kr1,ms1,us1} that 
$\zeta^\prime_q$ is a decreasing function of $q$ indicates that these
correlation functions exhibit multiscaling similar to that found in
turbulence. The quantity that is analogous to the energy dissipation rate
$\epsilon \sim (\partial u/\partial x)^2$ is the nearest-neighbor height
difference $s({\bf x},t) \equiv |h({\bf x}^\prime,t) - h({\bf x},t)|$ where 
${\bf x}^\prime$ is a nearest neighbor of the lattice point $\bf x$.
In analogy with Eq.(\ref{eqn3}), different
moments of $s$ scale with different powers of the correlation
length $\xi$:
\begin{equation}
\sigma_q(t) \equiv \langle [s({\bf x},t)]^q \rangle^{1/q} \approx
\xi(t)^{\alpha_q},
\label{eqn7} 
\end{equation}
where $\alpha_q$ are found 
\cite{kr1,ms1,us1} to increase with increasing $q$. 

In experimental and numerical studies of turbulence at relatively
low Reynolds numbers, the range of
values of $r$ for which the power-law scaling of Eq.(\ref{eqn2}) is observed
is often very small 
and sometimes non-existent. ESS in turbulence refers to the fact \cite{ess1} 
that the size of the scaling region is significantly enhanced if 
$\log |D_q(r)|, \,q \ne 3$ is plotted against $\log |D_3(r)|$.
Since the result $|D_3(r)| \propto r$
in the inertial range is exact for the Navier-Stokes equation underlying 
fluid turbulence and log-log plots of $|D_q(r)|$ against $|D_3(r)|$ show
linear behavior over a substantially larger range than
log-log plots of $|D_q(r)|$ against $r$,
ESS plots provide a convenient
way of determining the values of the multiscaling exponents $\zeta_q$.
ESS has been used extensively \cite{fn2} during the
last few years \cite{ess3,ess4,ess5} to analyze the data obtained from a 
variety of experiments and simulations. 
Substantial enhancement of the scaling region has also been observed
in log-log plots of $|G_p(r)|$ vs $|G_q(r)|$ where $p$ and $q$ are any two
unequal positive integers. 
The observation of ESS 
implies that the self-similarity of the velocity field extends
beyond the conventionally defined inertial range.

We have found that the correlation functions of height fluctuations in 
the Das Sarma-Tamborenea (DT) model \cite{dt} of epitaxial growth in one
and two dimensions 
exhibit properties that are very similar
to the ESS described above. 
Krug's original work \cite{kr1} on intermittent
multiscaling in epitaxial growth was based on the one-dimensional (1d)
DT model. 
In this model, atoms are
deposited randomly on a flat substrate under solid-on-solid 
condition. If a deposited atom has at least one lateral
neighbor, it stays at that site. Otherwise, the atom moves
to a nearest-neighbor lateral site if it can increase 
the number of lateral neighbors by doing so. If more than one such sites
are available, then the atom moves to any one of them with equal
probability. If no such 
nearest-neighbor site is available, the atom stays at the deposition
site. These rules are illustrated in the inset of Fig.1.
\begin{figure}
 \vbox to 5.5cm {\vss\hbox to 6cm
 {\hss\
   {\includegraphics{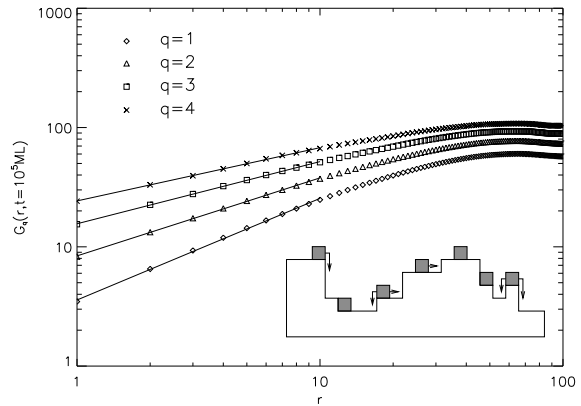}
   }
  \hss}
 }
\caption{The height-difference correlation functions $G_q(r),\, q=1-4$
at
time $t = 10^5$ as functions of the separation $r$ for the 1d DT model.
The
straight lines are best power-law fits of the data for $r \leq 10$.
Inset:
Rules of the DT growth model. Random deposition on the substrate
is followed by nearest-neighbor diffusion if it is allowed by the local
coordination, as indicated by the arrows.}
\end{figure}
The DT model
has been studied extensively \cite{kr1,dt,ms1} by simulations. 

We first consider the scaling properties of the correlation
functions $G_q(r)$ defined in Eq.(\ref{eqn5}). Fig.1 shows the results for
$G_q(r), \,q = 1-4$ at time $t = 10^5$ (in units of number of layers deposited)
for the 1d DT model with size $L$ = 1000. 
The log-log
plots of $G_q(r)$ against $r$ clearly show linear behavior for 
small $r$, in
agreement with the 
power-law form of Eq.(\ref{eqn6}). The values of the exponents
$\zeta^\prime_q$ obtained by fitting straight lines to the initial linear
parts of the plots are: $\zeta^\prime_1 = 0.88 \pm 0.02$,
$\zeta^\prime_2 = 0.67 \pm 0.02$, $\zeta^\prime_3 = 0.54 \pm 0.02$, and
$\zeta^\prime_4 = 0.45 \pm 0.02$. The
size of the scaling region is, however, rather small (less
than one decade in $r$).
This is not surprising because the
value of the correlation length $\xi \propto t^{1/z}, \,z \simeq$ 4 
\cite{kr1,dt}, is 
the order of 15 at $t = 10^5$. 
The new and surprising result we find is that the data exhibit
nearly perfect power-law scaling over the entire range $1 \le r \le 100$ if
 $G_q(r), \, q\ne 1$ is plotted against $G_1(r)$ (see Fig.2). Since there is
\begin{figure}
 \vbox to 5.5cm {\vss\hbox to 6cm
 {\hss\
   {\includegraphics{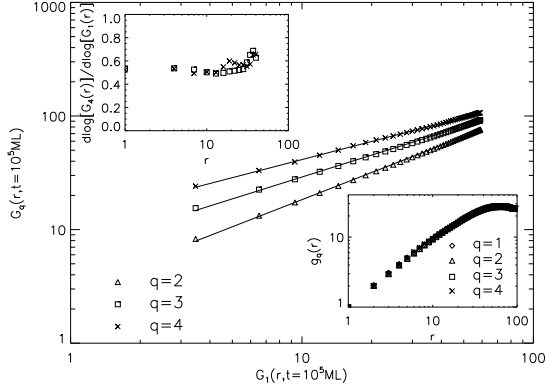}
   }
  \hss}
 }
\caption{ Log-log plots of $G_q(r), \, q=2-4$ of Fig.1 against $G_1(r)$.
The
straight lines are best fits to power laws. Upper inset: The local slope
$d\log[G_4(r)]/d\log[G_1(r)]$ of the $q = 4$ plot in the main
figure as
a function of $r$ (crosses), and same quantity for $L$ = 100
in the saturation regime (squares). Lower inset: The
scaling
functions $g_q(r),\,q=1-4$ (see text).}
\end{figure}
no ``special'' value of $q$ in the growth model (in the sense that $q = 3$
is special in turbulence), we have chosen to plot the correlation functions
for $q \ne 1$ against $G_1$ -- other choices lead to equally good scaling.
It is clear from Fig.2 that the range over which
scaling behavior is observed is increased by more than an order of magnitude
if one of the correlation functions is plotted against the other, rather than
plotting them as functions of $r$. This model, therefore, exhibits ESS that
is formally identical to that found in turbulence by Benzi 
{\it et al} \cite{ess1}. The extension of the scaling range is mostly
towards {\em large} scales, which is similar to
the behavior found \cite{glr} in turbulence. 

The quality of the power-law scaling in the $G_q,\,q \ne 1$ versus $G_1$ plots
is illustrated in the 
upper inset of Fig.2 for $q = 4$, where we have shown the
variation of the local derivative $d\log(G_4(r))/d\log(G_1(r))$ with $r$. The
observed variation is less than 10\%, 
indicating that the power-law relation, 
$G_q(r,t) \propto [G_1(r,t)]^{\theta_q}, \, q \ne 1$,
provides a very good description of the data over the full range of $r$. The
values of the ``relative'' exponents $\theta_q$, obtained from straight-line
fits to the plots of Fig.2, are: $\theta_2 = 0.79 \pm 0.02$, 
$\theta_3 = 0.64 \pm
0.02$, and $\theta_4 = 0.53 \pm 0.02$. These values
are consistent with the expected result, $\theta_q = 
\zeta_q^\prime/\zeta_1^\prime$.
The upper inset of Fig.2 also shows the local slope of a log-log plot
of $G_4(r)$ versus $G_1(r)$ obtained in the saturation state of samples with
$L$ = 100. The two sets of 
results are nearly indistiguishable, showing that the values of the exponents
$\theta_q$ obtained from ESS plots are 
not sensitive to details such as sample size 
and the length of the simulation. Similar ``universality'' (i.e.
insensitivity of the values of the relative exponents to 
details such as the value of the
Reynolds number and the flow geometry) has been observed in turbulence 
\cite{ess4,ess5}.

The observation of ESS implies a specific relation among the 
scaling functions \cite{kr1} that
describe the behavior of the correlation functions
$G_q(r,t)$. The dependence of $G_q(r,t)$ on $r$ and $t$ in 
the regime $\xi \approx t^{1/z} \ll L$ is expected to 
be described by the scaling form \cite{kr1}
\begin{equation}
G_q(r,t) = t^{\alpha_q/z} r^{\zeta^\prime_q}f_q(r/t^{1/z}).
\label{eqn9}
\end{equation}
The occurrence of ESS is possible only if the scaling functions $f_q$
are related to one another by 
$f_q(r/t^{1/z}) = C_q(t) [f(r/t^{1/z})]^{\zeta^\prime_q}$.
This relation, when combined with Eq.(\ref{eqn9}), leads to the 
prediction that plots of the quantity $g_q(r,t) \equiv [G_q(r,t)/G_q(r=1,t)]^
{1/\zeta^\prime_q}$ against $r$ should coincide for all values of $q$. The
lower inset of Fig.2 shows plots of $g_q(r)$ (obtained from the
data shown in Fig.1, using the values 0.870, 0.673, 0.540 and 0.452
for $\zeta^\prime_q, \, q=1-4$, respectively) against $r$ for $q = 1-4$. 
The data for different $q$ collapse nicely to the same curve, confirming the
occurrence of ESS in this system.
A similar description of ESS in turbulence in terms of a 
$q$-independent scaling function is provided in Ref.\cite{ess3}.

The behavior described above is also found in the 2d DT model, as shown in
Fig.3, where we have plotted $G_q(r), \,q = 1-4$ against $r$ (inset),
\begin{figure}
 \vbox to 5.5cm {\vss\hbox to 6cm
 {\hss\
   {\includegraphics{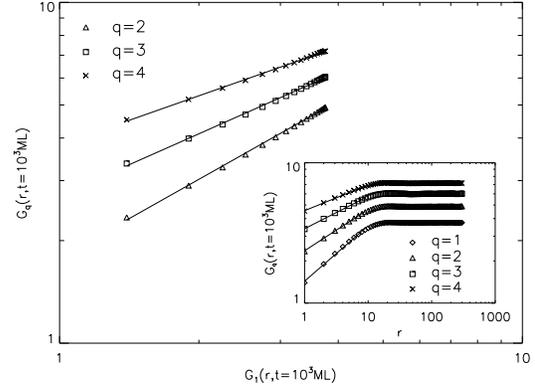}
   }
  \hss}
 }
\caption{
Log-log plots of $G_q(r), \, q=2-4$ against $G_1(r)$ for the 2d DT model
at
time $t = 10^3$. The
straight lines are best fits to power laws.
Inset: Log-log plots of $G_q(r),\, q=1-4$
as functions of the separation $r$ for the same data.}
\end{figure}
and $G_q(r), \, q=2-4$ against $G_1(r)$ (main part) on log-log scales. 
The data shown correspond to \cite{ms1} $t = 10^3$ for
500$\times$500 samples. 
We again
find an extension of the scaling region by more than one decade.
The values of the exponents obtained from fits to the data are: $\theta_2 = 
0.77 \pm 0.02$, $\theta_3 = 0.60 \pm 0.02$, $\theta_4 = 0.48 \pm 0.02$.
As before, the relation $\theta_q = \zeta^\prime_q/\zeta^\prime_1$ is satisfied
within the error bars.

We have also found ESS in the dependence of the quantities $\sigma_q$ defined
in Eq.(\ref{eqn7}) on time $t$ (in the {\it growth}
regime) or the sample size $L$
(in the {\it saturation} regime).
\begin{figure}
 \vbox to 5.5cm {\vss\hbox to 6cm
 {\hss\
   {\includegraphics{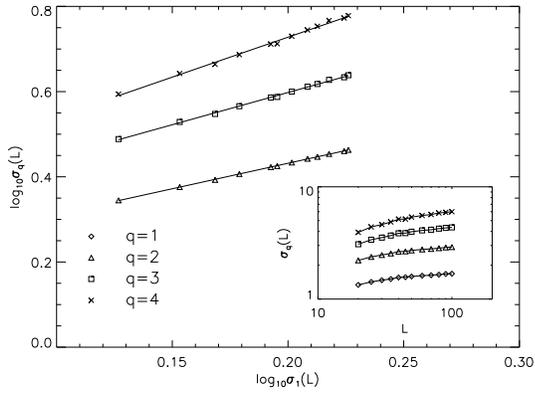}
   }
  \hss}
 }
\caption{ Log-log plots of $\sigma_q(L), \, q=2-4$
against
$\sigma_1(L)$ for the 2d DT model at saturation. The
straight lines are best fits to power laws.
Inset: Log-log plots of $\sigma_q,\,q=1-4$
as functions of the system size $L$ for
the 2d DT model at saturation.}
\end{figure}
The inset of Fig.4
shows our data for the dependence of $\sigma_q,\, q=1-4$ 
at saturation on $L$ in the 2d DT model.
The log-log plots clearly show a downward curvature (similar curvature
is observed \cite{ms1} in log-log plots of $\sigma_q$ versus $t$ in the
growth regime) that becomes more 
pronounced as $L$ is increased. 
In contrast, the ESS plots of $\sigma_q,\, q=2-4$ against
$\sigma_1$, shown in Fig.4, do not exhibit any such curvature. 
The values
of the ``relative'' exponents $\psi_q$, defined by $\sigma_q \propto
(\sigma_1)^{\psi_q}$, 
obtained from straight-line fits to the the ESS plots are: $\psi_2 \simeq 1.2$,
$\psi_3 \simeq 1.5$, $\psi_4 \simeq 1.9$.

The results shown in Fig.4 and
qualitatively similar behavior found in the 
1d DT model \cite{kr1,ms1}
clearly indicate that the DT model {\em does not} exhibit true 
asymptotic multiscaling. 
This conclusion is corroborated by recent studies
\cite{us1} which suggest that the approximate multiscaling observed in
these models is a 
non-universal and extremely 
slow transient or crossover arising from a non-linear
instability in the discretized version of the underlying continuum 
growth equation. The ESS found in this paper
shows that the slow crossover responsible for the approximate multiscaling
behavior affects correlation functions of different 
order in exactly the same way. In this picture, the 
multiscaling exponents $\zeta^\prime_q$
and $\alpha_q$ are, at best, {\em effective} ones describing the behavior
over a limited range of length and time scales. The occurrence of ESS 
implies that ``relative'' exponents such as 
$\theta_q$ and $\psi_q$ are, in some 
sense, more fundamental than the 
``absolute'' ones, $\zeta^\prime_q$ and $\alpha_q$.
While it would be premature to suggest that 
a similar transient description applies also to multiscaling in
turbulence, we note that the possibility that the intermittency corrections
$\delta \zeta_q$ are finite Reynolds number effects which would vanish in
the $Re \rightarrow \infty$ 
asymptotic limit has received considerable attention \cite{lp}
in the recent literature.
Also, there is some numerical evidence \cite{ls} indicating
that the ``relative''
exponents are more universal than the ``absolute'' ones in 
turbulence. 

Our finding that ESS may occur in problems (i.e. in the kinetic surface
roughening of the DT model) very different from the fully developed
homogeneous turbulence problem (where ESS was originally discovered
\cite{ess1}) is potentially significant, and may eventually provide a
clue to its understanding. Currently the ESS phenomenon (both in surface
growth and turbulence) remains an interesting empirical fact without any
rigorous theoretical understanding. It has recently been shown \cite{us1}
that multiscaling in the DT model shares a substantial common
phenomenology with that in fully developed turbulence, and the
intermittent multiscaling behavior in these very different problems may
arise \cite{us1} from the existence of an infinite number of marginal
operators and an associated near-singularity in both problems. We
speculate that the eventual theoretical understanding of ESS will depend
on a more detailed understanding of the roles that the infinite number
of relevant operators and the near-singularities play in homogeneous
turbulence and kinetic surface roughening.
It would obviously be very interesting to 
explore further 
the analogy between kinetically rough interfaces and fluid turbulence.  

We are grateful to Prof. Rahul Pandit for educating us on ESS in turbulence.
This work was supported in part by US-ONR and 
the Supercomputer 
Education and Research Center of Indian Institute of Science. 
 
%\end{document}

\begin{references}
\bibitem{fn1} In contrast with Eq.(\ref{eqn1}), the
$q$-th root is included in the definition of $G_q(r,t)$ in
order to maintain conformity with the definitions used in earlier work 
\cite{kr1,ms1,us1}.
\end{references}
\end{document}